\documentclass{article}

\usepackage[T1]{fontenc}    
\usepackage[english,frenchb]{babel}
\usepackage[space]{grffile}
\usepackage[utf8]{inputenc} 
\usepackage{algorithm}
\usepackage{algpseudocode}
\usepackage{amsfonts}       
\usepackage{amsmath}
\usepackage{amssymb}
\usepackage{amsthm}
\usepackage{arxiv}
\usepackage{babel}
\usepackage{booktabs}       
\usepackage{color}
\usepackage{graphicx}
\usepackage{hyperref}       
\usepackage{latexsym}
\usepackage{lipsum}		
\usepackage{microtype}      
\usepackage{multirow}
\usepackage{nicefrac}       
\usepackage{times}
\usepackage{url}            
\usepackage{verbatim}

\makeatletter

\providecommand{\tabularnewline}{\\}
\newcommand{\lyxdot}{.}

\floatstyle{ruled}
\newfloat{algorithm}{tbp}{loa}
\providecommand{\algorithmname}{Algorithm}
\floatname{algorithm}{\protect\algorithmname}

\@ifundefined{date}{}{\date{}}

\title{Plackett-Luce Model for Learning-to-Rank Task}
\author{
  Tian Xia \\
  Wright State University \\
  \texttt{SummerRainET2008@gmail.com} \\
  \And
  Shaodan Zhai \\
  Wright State University \\
  \texttt{ShaodanZhai@gmail.com} \\
  \And
  Shaojun Wang \\
  Wright State University \\
  \texttt{SWang.USA@gmail.com} \\
}

\begin{document}
  \maketitle

  \begin{abstract}
    List-wise based learning to rank methods are generally supposed to
    have better performance than point- and pair-wise based. However,
    in real-world applications, state-of-the-art systems are not from
    list-wise based camp. 
    In this paper, we propose a new non-linear algorithm in the list-wise
    based framework called ListMLE, which uses the Plackett-Luce (PL)
    loss. Our experiments are conducted on the two largest publicly available
    real-world datasets, Yahoo challenge 2010 and Microsoft 30K. This
    is the first time in the single model level for a list-wise based
    system to match or overpass state-of-the-art systems in real-world
    datasets. 
  \end{abstract}

  \section{Introduction}

  The learning to rank task arises from real-world applications such
  as Google, Yahoo, and other search engines. A ranking system returns
  a set of documents and ranks them by their relevance to the query
  from a user. 

  Learning to rank techniques are influencing traditional natural language
  processing applications, such as model parameter training \cite{tuning-as-ranking},
  and non-linear feature extraction \cite{non-linear-features1,non-linear-features2}. 

  Generally, ranking models fall into three methodologies based on how
  they model basic ranking objects.\emph{ This definition would not
  be affected by how to utilize features}, e.g., linear and non-linear
  features. 

  The first methodology, \textbf{point-wise based}, breaks relationship
  between documents related to different queries \cite{cossock2006subset,crammer2001pranking,friedman2001greedy,li2007mcrank},
  then uses traditional machine learning regression and classification
  techniques for training. For example, MART \cite{friedman2001greedy}
  uses the regression tree technique to fit model outputs to their relevance
  scores; McRank \cite{li2007mcrank} converts the rank procedure as
  a multi-class classification.

  The second methodology, \textbf{pair-wise based}, considers the relationship
  among documents related to the same query \cite{cohen1999learning,freund2003efficient,hazan2010direct,herbrich1999support,joachims2002optimizing,quoc2007learning,rudin2009p,tsai2007frank,wu2010adapting},
  then adopts mature classification techniques to minimize the inversion
  number of documents by considering document pairs. For example, RankBoost
  \cite{freund2003efficient} plugs the exponential loss of document
  pairs into a framework of Adaboost; RankSVM \cite{herbrich1999support,joachims2002optimizing}
  uses SVM to perform a binary classification on the document pairs;
  LambdaRank \cite{quoc2007learning} and LambdaMART \cite{wu2010adapting}
  take into account the influence of a correctly classified document
  pair to the objective measures, and achieve a big success.

  The third methodology, \textbf{list-wise based}, treats a permutation
  of a set of documents as a basic unit, and builds loss functions on
  them \cite{cao2007learning,metzler2007linear,ravikumar2011ndcg,tan2013direct,xia2009statistical,xia2008listwise,xu2007adarank,xu2008directly}.
  Because exact losses of performance measures are step-wise, non-differentiable
  as well as non-convex with respect to model parameters, most work
  in this methodology resort to suitable surrogate functions. These
  surrogate functions are either not directly related to ranking performance
  measures \cite{cao2007learning,qin2008query,xia2009statistical,xia2008listwise},
  or just continuous and differentiable approximation bounds of ranking
  measures \cite{chakrabarti2008structured,chapelle2010gradient,le2007direct,qin2010general,taylor2008softrank,valizadegan2009learning,wu2010adapting,xu2008directly,xu2007adarank,yue2007support}.
  To further decrease the gap between optimization objectives and performance
  measures, some work attempt to directly optimize objective measures
  and show promising results. For example, in \cite{metzler2007linear,tan2013direct},
  the authors use a coordinate ascent framework to directly optimize
  performance measures, and DirectRank in \cite{tan2013direct} is much
  faster in practice. However, both their work still can not match the
  state-of-the-art systems in large data sets when decision trees are
  used \footnote{Tan et al. \cite{tan2013direct} use a mixed strategy, which borrows
  boosted trees generated from MART, to compete with LambdaMART. Their
  strategy should be treated as a system combination technique rather
  than a single ranking model.}.

  Our work utilizes an elegant list-wise surrogate function called Plackett-Luce
  (PL) loss, which was first proposed in 1975 \cite{plackett1975analysis}
  for horse gambling. Cao et al. \cite{cao2007learning} introduce it
  to the learning to rank task by using it to model the probabilistic
  distribution of a set of documents given a query, where the training
  is conducted by minimizing the KL distance between the probability
  distribution for the ranking model and that for the ground truth.
  Later Xia et al.\cite{xia2009statistical,xia2008listwise} provide
  a model called ListMLE, which instead maximizes the likelihood of
  ground-truth permutations defined in the PL loss. ListMLE could be
  viewed as \emph{a general framework} to utilize linear and non-linear
  features, however, as its non-linear system has not been developed,
  we refer to $l$-ListMLE as its linear version hereafter. Because
  public large-scale datasets were not available until 2010, many properties
  of the PL loss are not revealed in $l$-ListMLE. Even though $l$-ListMLE
  performs pretty well on some datasets, it is rather unstable in many
  other cases, especially when compared with direct optimization based
  models, e.g. DirectRank \cite{tan2013direct} and LambdaRank \cite{quoc2007learning}.
  Although not necessarily the best, DirectRank and LambdaRank often
  show reasonable good performance, while $l$-ListMLE under some circumstances
  performs far more poorly than average performance. For example, on
  the Microsoft 30K data, the largest publicly available real world
  dataset, $l$-ListMLE is approximately 7.6 points worse than the coordinate
  ascent based method \cite{metzler2007linear} in terms of NDCG scores.
  Although, Xia et al. \cite{xia2009statistical} further proved the
  PL loss is consistent with NDCG@K under certain assumptions, it is
  not guaranteed to achieve a reasonable performance on practical applications
  that use data sets with limited size, and the unstable performance
  behavior greatly limits wide spread real-world applications for the
  ListMLE model.

  Understanding why the PL loss fails in some datasets is important
  to design more effective algorithms, thus we conduct experiments to
  analyze these datasets, and figure out one principle as the condition
  for the PL loss, which states that as compared to average document
  number per query, the number of features should be large enough. Therefore
  in order to gain better performance, we have to use more features
  for PL loss. There are several ways to enrich features of datasets:
  \emph{kernel mapping}, \emph{neural network mapping}, and \emph{gradient
  boosting}. We select the gradient boosting with decision trees as
  weak rankers in this work due to the convenient comparison with LambdaMART,
  and leave the others for further work. A merit of the PL loss is its
  concise formula to compute functional gradients, Eqn. (\ref{eq:numerator}),
  which results in our ranking system, called PLRank. 

  As suggested in \cite{chapelle2011yahoo}, real-world datasets are
  closer to the scenario of search engine applications and have much
  smaller fluctuations in terms of performance. We conduct experiments
  on two publicly released real-world datasets. As far as we know, these
  datasets are larger than any used in previous research papers, except
  \cite{wu2010adapting} \footnote{They adopted a larger but proprietary one}.
  To compare with other list-wise based methods, we also extend three
  extra consistent list-wise surrogate functions in \cite{ravikumar2011ndcg}
  in the gradient boosting framework. We find that PLRank not only maintains
  the merits of the PL loss, but also greatly alleviates the instability
  problem of $l$-ListMLE. PLRank has the same time complexity with
  LambdaMART, and is $M$ times as fast as McRank \footnote{$M$ is the number of different relevance scores in measuring a document.}.

  \section{Background}

  \subsection{Basic Notations}

  Given a set of queries $\mathbf{Q}=\{q_{1},\ldots,q_{|\mathbf{Q}|}\}$,
  each query $q_{i}$ is associated with a set of candidate relevant
  documents $\mathbf{D}_{i}=\{d_{1}^{i},\ldots,d_{|\mathbf{D}_{i}|}^{i}\}$
  and a corresponding vector of relevance scores $\mathbf{r}_{i}=\{r_{1}^{i},\ldots,r_{|\mathbf{D}_{i}|}^{i}\}$
  for each $\mathbf{D}_{i}$. The relevance score is usually an integer,
  and greater value means more related for the document to the query.
  An $M$-dimensional feature vector $\mathbf{h}(d)=[h_{1}(d|q),\ldots,h_{M}(d|q)]^{T}$
  is created for each query-document pair, where $h_{t}(\cdot)$s are
  predefined real-value feature functions.

  A ranking function $f$ scores each query-document pair, and returns
  sorted documents associated with the same query. Since these documents
  have a fixed ground truth rank, our goal is to learn an optimal ranking
  function returning results as close to the ground truth rank as possible.

  Generally, ranking functions use only linear information of original
  features $\mathbf{h}(d|q)$ or their nonlinear information. The linear
  form is as $f(d|q)=\boldsymbol{w}^{T}\cdot\mathbf{h}(d)$, where $\mathbf{\boldsymbol{w}}=[w_{1},\ldots,w_{M}]^{T}\in\mathcal{R}^{M}$
  is the model parameter. The nonlinear form often adopts regression
  trees, kernel technique, and neural network.

  Several measures have been used to quantify the quality of a rank,
  such as NDCG@K, ERR, MAP etc. In this paper, we use the most popular
  NDCG@K and ERR \cite{chapelle2011yahoo} as the performance measures.

  \subsection{Gradient Boosting and Regression Tree}

  We review gradient boosting \cite{friedman2001greedy} as a general
  framework for function approximation using regression trees as the
  weak learners, which has been the most successful approach for learning
  to rank models.

  Gradient boosting iteratively finds an additive predictor $f(\cdot)\in\mathcal{H}$
  that minimizes a loss function $\mathcal{L}$. At the $t$th iteration,
  a new weak learner $g_{t}(\cdot)$ is selected to be added to current
  predictor $f_{t}(\cdot)$ to construct a new predictor, 
  \begin{equation}
    f_{t+1}(\cdot)=f_{t}(\cdot)+\alpha g_{t}(\cdot)\label{eq:update}
  \end{equation}
  where $\alpha$ is the learning rate.

  In gradient boosting, according to the following squared loss, $g_{t}(\cdot)$
  is chosen as the one most parallel to the pseudo-response $-\frac{\partial\mathcal{L}}{\partial f_{t}(\cdot)}$,
  which is negative derivative of the loss function in functional space.
  \begin{equation}
    g_{t}(\cdot)=\arg\min_{g\in\mathcal{H}}\parallel-\frac{\partial\mathcal{L}}{\partial f_{t}(\cdot)}-g(\cdot)\parallel_{2}^{2}\label{eq:MSE}
  \end{equation}

  To fit a regression tree, the data in each internal tree node is greedily
  splitted into two parts by minimizing Eqn. (\ref{eq:MSE}), and this
  procedure recursively iterates until a predefined condition is satisfied.
  This tree construction procedure is applicable for any differentiable
  loss function. The complexity of a regression tree is usually controlled
  by the tree height or leaf number. In learning to rank, the latter
  is more flexible, thus is adopted in this work by default.

  \section{Plackett-Luce Loss for Learning to Rank}

  The Plackett-Luce model was first proposed by Plackett \cite{plackett1975analysis}
  to predict the ranks of horses in gambling. Consider a horse racing
  game with five horses. Suppose a probability distribution $\mathcal{P}$
  on their abilities to win a race, then a rank of these horses can
  be understood as a generative procedure. Suppose we want to know the
  probability of a top3 rank $\overline{2,3,5}$. The result can be
  computed as follows:

  Being the champion for the 2nd horse, the probability is $p_{2}$
  among five candidates. Being the runner-up for the 3rd horse, the
  probability $p_{3}$ has to be normalized among the remaining four
  horses, which leads to $p_{3}/(p_{1}+p_{3}+p_{4}+p_{5})$. Being the
  third winner for the 5th horse, its probability among the remaining
  three horses becomes $p_{5}/(p_{1}+p_{4}+p_{5})$. So the probability
  of the rank $\overline{2,3,5}$ is their product. It is not difficult
  to see that the most likely rank is all horses are ranked by their
  winning probability in a descending order.

  The key idea for the Plackett-Luce model is the choice in the $i$th
  position in a rank $\pi$ only depends on the candidates not chosen
  at previous positions.

  \subsection{Plackett-Luce Loss with Linear Features}

  In learning to rank, each training sample has been labeled with a
  relevance score, so the ground-truth permutation of documents related
  to the $i$th query can be easily obtained and denoted as $\pi_{i}$,
  where $\pi_{i}(j)$ denotes the index of the document in the $j$th
  position of the ground-truth permutation. We note that $\pi_{i}$
  is not obligatory to be a full rank, as we may only care about the
  top $K$ documents.

  Consider a ranking function with linear features, the probability
  of a set of candidate relevant documents $\mathbf{D}_{i}$ associated
  with a query $q_{i}$ is defined as 
  \begin{equation}
    p(d_{e}^{i})=\frac{\exp\{\mathbf{h}(d_{e}^{i})^{T}\cdot\boldsymbol{w}\}}{\sum_{d\in\mathbf{D}_{i}}\exp\{\mathbf{h}(d)^{T}\cdot\boldsymbol{w}\}}\label{eq:prob}
  \end{equation}
  The probability of the Plackett-Luce model to generate a rank $\pi_{i}$
  is given as 
  \begin{eqnarray}
    p(\mathbf{\pi}_{i},\boldsymbol{w})=\prod_{j=1}^{|\pi_{i}|}p(d_{\pi_{i}(j)}^{i}|C_{i,j}) &  & p(d_{e}^{i}|C_{i,j})=\frac{p(d_{e}^{i})}{\sum_{d\in C_{i,j}}p(d)}\label{eq: full-formula}
  \end{eqnarray}
  where $C_{i,j}=\mathbf{D}_{i}-\{d_{\pi_{i}(1)}^{i},\ldots d_{\pi_{i}(j-1)}^{i}\}$.

  The training objective is to maximize the log-likelihood of all expected
  ranks over all queries and retrieved documents with corresponding
  ranks in the training data with a zero-mean and unit-variance Gaussian
  prior parameterized by $\boldsymbol{w}$. 
  \begin{equation}
    \mathcal{L}=\log\{\prod_{i}p(\pi_{i},\boldsymbol{w})\}-\frac{1}{2}\boldsymbol{w}^{T}\boldsymbol{w}\label{eq:linear-obj}
  \end{equation}
  {\footnotesize{} }{\footnotesize\par}

  The gradient can be calculated as follows, 
  \[
  \frac{\partial\mathcal{L}}{\partial\boldsymbol{w}}=\sum_{i}\sum_{j}\{\mathbf{h}(d_{\pi_{i}(j)}^{i})-\sum_{d\in C_{i,j}}(\mathbf{h}(d)\cdot p(d|C_{i,j}))\}-\boldsymbol{w}
  \]
  Since the log-likelihood function is smooth, differentiable, and
  concave with the weight vector $\boldsymbol{w}$, global optimum guarantee
  is satisfied.

  \subsection{Plackett-Luce Loss with Regression Trees}

  In this paper, we build ensemble regression trees for the Plackett-Luce
  loss in the gradient boosting framework, Alg. \ref{alg:PL1} summarizes
  the main procedure. We first describe how to compute the pseudo response
  and output value for fitting a regression tree, and then we provide
  more analysis for this new model.

  At the $t$th iteration, all fitted regression trees constitute the
  current predictor $f_{t}(\cdot)$, and the Eqn. (\ref{eq:prob}) can
  be rewritten as 
  \begin{equation}
    p(d_{e}^{i})=\frac{\exp\{f_{t}(d_{e}^{i})\}}{\sum_{k=1}^{|\mathbf{D}_{i}|}\exp\{f_{t}(d_{k}^{i})\}}\label{eq:prob1}
  \end{equation}

  We limit $|\pi|=K$, and adopt Eqn. (\ref{eq:linear-obj}) without
  a normalization as our objective \footnote{The model complexity of regression trees is often controlled by the
  learning rate $\alpha$, different from the normalization factor used
  in a linear model.}. Plugging Eqn. (\ref{eq:prob1}) into Eqn. (\ref{eq:linear-obj}),
  and taking derivative with respect to $f_{t}(\cdot)$, we obtain 
  \begin{equation}
    \mathcal{L}'(f_{t}(d))=\textrm{I}(d\in\textrm{top}K\textrm{ ground-truth})-\sum_{C\,s.t.\,d\in C}p(d|C)\label{eq:first-order}
  \end{equation}
  where $\textrm{I}(\cdot)$ denotes the indicator function. When $\textrm{I}(\cdot)$
  returns 0 for the current document, the size of $\{C\}$ equals $K$,
  otherwise it is smaller.

  We follow Eqn. (\ref{eq:MSE}) to fit a regression tree $g_{t}(\cdot)$.
  Denotes the documents falling in the leaf $U$ as $U_{d}$. We set
  the output of the leaf $U$ as $g_{t}(d\in U_{d})=-v$, and $v$ is
  optimized independently from other leaves. Following Eqn. (\ref{eq:update}),
  we construct $f_{t+1}(\cdot)$ for documents in $U_{d}$.

  We adjust $v$ to maximize the log-likelihood $\mathcal{L}$. Thus
  $\mathcal{L}$ has been reinterpreted as a function of $v$. We rewrite
  Eqn. (\ref{eq:prob1}) as 
  \begin{equation}
    p(d_{e}^{i})=\frac{\exp\{f_{t}(d_{e}^{i})-\textrm{I}(d_{e}^{i}\in U_{d})\cdot\alpha v\}}{\sum_{k=1}^{|\mathbf{D}_{i}|}\exp\{f_{t}(d_{k}^{i})-\textrm{I}(d_{e}^{i}\in U_{d})\cdot\alpha v)\}}
  \end{equation}
  By the Newton method, we have $v=\frac{\mathcal{L}'(v=0)}{\mathcal{L}''(v=0)}$,
  where

  \begin{eqnarray}
    \mathcal{L}'(v=0)=\sum_{d\in U_{d}}\mathcal{L}'(f_{t}(d)), & \mathcal{L}''(v=0)=\sum_{C}p'\cdot(p'-1), & p'=\sum_{d\in U_{d}\cap C}p(d|C)\label{eq:numerator}
  \end{eqnarray}

  \begin{algorithm}[h!]
    \begin{algorithmic}[1]

      \Require Documents $\mathbf{D}=\{\mathbf{D}_{1},\mathbf{D}_{2},\ldots\}$;
      $K$ defines top$K$ documents of a ground-truth rank; $T$ defines
      regression tree number; $L$ defines leaf number; $\alpha$ defines
      learning rate.

      \State $f_{1}(\cdot)\leftarrow$ BackGroundModel($\cdot$) \Comment{Initialization
      for model adaptation. None by default.}

      \For {$\mathbf{D}_{i}$ in $\mathbf{D}$} \State {Randomly shuffle
      $\mathbf{D}_{i}$}

      \State Sort $\mathbf{D}_{i}$ by relevances. \Comment{We could
      build several ground-truth permutations.}

      \EndFor

      \For {$t=1$ to $T$} \State $Resp(d\in\bigcup\mathbf{D_{i}})\leftarrow-\mathcal{L}'(f_{t}(d))$
      \Comment{Compute pseudo response following Equ. \ref{eq:first-order}.}

      \State Fit a $L$-leaf tree $g_{t}$ on $Resp$. \Comment{By Eqn.
      (\ref{eq:MSE}) by default.}

      \For {leaf $U$ in $g_{t}$} 

      \State $v\leftarrow\mathcal{L}'(v=0)/\mathcal{L}''(v=0)$ \Comment{Set
      output of current leaf by Eqn. (\ref{eq:numerator})} 

      \State $g_{t}(d\in U_{d})\leftarrow-v$ 

      \EndFor

      $f_{t+1}\leftarrow f_{t}+\alpha g_{t}$ \Comment{Eqn. (\ref{eq:update})} 

      \EndFor

      \Return $f_{T+1}$ \caption{PLRank}

      \label{alg:PL1}

    \end{algorithmic} 
  \end{algorithm}
  To clarify this procedure, we take one query with four related documents
  as an example. Suppose the four documents $d_{1},d_{2},d_{3},d_{4}$
  are sorted in a descending order with their relevance scores. In an
  other word, the ground-truth permutation is $d_{1},d_{2},d_{3},d_{4}$.
  Let their scores after some iterations, from current predictor $f_{t}(\cdot)$,
  be $s_{1},s_{2},s_{3},s_{4}$ respectively for abbreviation. Considering
  the top $2$ documents of the ground-truth permutation, the log-likelihood
  is 
  \begin{align*}
    \mathcal{L}= & s_{1}-\log\{\exp s_{1}+\exp s_{2}+\exp s_{3}+\exp s_{4}\}+s_{2}-\log\{\exp s_{2}+\exp s_{3}+\exp s_{4}\}
  \end{align*}
  Taking derivatives with respect to their scores, we obtain 
  \[
  \begin{array}{cl}
    \mathcal{L}'(s_{1})=1-p(s_{1}|s_{1},s_{2},s_{3},s_{4}), & \mathcal{L}'(s_{2})=1-p(s_{2}|s_{1},s_{2},s_{3},s_{4})-p(s_{2}|s_{2},s_{3},s_{4})\\
    \mathcal{L}'(s_{3})=0-p(s_{3}|s_{1},s_{2},s_{3},s_{4})-p(s_{3}|s_{2},s_{3},s_{4}) & \mathcal{L}'(s_{4})=0-p(s_{4}|s_{1},s_{2},s_{3},s_{4})-p(s_{4}|s_{2},s_{3},s_{4})
  \end{array}
  \]
  In this toy example, the samples $s_{3},s_{4}$ have $K=2$ contextual
  probabilities.

  Suppose $s_{1}$,$s_{3}$ fall into the same leaf of a regression
  tree, then 
  \[
  \begin{array}{cl}
    \mathcal{L}'(v=0) & =1-p(s_{1}|C_{1})+0-\{p(s_{3}|C_{1})+p(s_{3}|C_{2})\}\\
    \mathcal{L}''(v=0) & =(p(s_{1}|C_{1})+p(s_{3}|C_{1}))\cdot(p(s_{1}|C_{1})+p(s_{3}|C_{1})-1)+p(s_{3}|C_{2})\cdot(p(s_{3}|C_{2})-1)
  \end{array}
  \]
  where $C_{1}=\{s_{1},s_{2},s_{3},s_{4}\}$, $C_{2}=\{s_{2},s_{3},s_{4}\}$.

  In the following, we describe more details of Alg. \ref{alg:PL1}
  that relate to initialization of models (line 1), selection of ground-truth
  permutation (line 3-4).

  \subsubsection{Initialization of Models}

  As a statistical model is sensitive to data genres, a trivial yet
  effective way is to use more data for training. Borrowing the idea
  from adaptive LambdaMART \cite{wu2010adapting}, our model could also
  first train a background model on plenty of general genre data. Then
  we assign the resulting model to initialize our Alg \ref{alg:PL1}
  (line 1), and continue to train our model using on objective genre
  data. In this paper, we are not focusing on the adaptation experiments,
  and we initialize to zero.

  \subsubsection{Selection of Ground-Truth Permutations\label{subsec:Selection-of-Ground-Truth}}

  In learning to rank, as the relevance scores are scattered among limited
  integers, e.g., 0 to 10 inclusively, there are many ties in the scores,
  this would impact the determination of ideal permutations and our
  training objective. 

  We consider \emph{multiple ground-truth permutations} (looping lines
  2-5 in Alg. \ref{alg:PL1}). Let toy documents be $d_{1},d_{2},d_{3},d_{4}$
  with relevance scores $4,0,4,4$, and considering top $4$ ground-truth
  documents. As the number of all permutation possibilities is huge,
  we randomly select several ground-truth ranks and store them compactly
  in terms of data structure. For instance, the ground truth permutation
  $d_{1},d_{3},d_{4},d_{2}$ consists of three contextual terms, $C_{1}=\{d_{1},d_{2},d_{3},d_{4}\}$,
  $C_{2}=\{d_{2},d_{3},d_{4}\}$, $C_{3}=\{d_{2},d_{4}\}$, while adding
  a second permutation $d_{1},d_{4},d_{3},d_{2}$ leads to merely one
  extra term $C_{4}=\{d_{2},d_{3}\}$, rather than new three terms.
  The statistics about this issue are in Table \ref{tbe:balance}. We
  use PLRank(obj=$num$) to denote different number of objectives.

  \subsection{Training with Plackett-Luce Loss}

  Regarding linear features, Xia et al. \cite{xia2009statistical,xia2008listwise}
  adopt a neural network to maximize the log-likelihood of expected
  ranks. The neural network works well in small datasets, e.g. LETOR,
  while it also requires suitable settings on hidden layer structure
  and the number of hidden neurons. 

  As our experiments are conducted on real-world datasets, we instead
  use L-BFGS \cite{byrd1995limited} for parameter tuning to gain faster
  convergence speed. It is observed that overfitting often occurs in
  small data sets, while in large datasets the the log-likelihood correlates
  with ranking measures very well.

  Regarding non-linear features, kernel technique could map them into
  a linear form in a high dimensional space, and then the neural network
  based training in Xia et al.'s work or LBFGS are applicable, provided
  that the new dimension is acceptable in practice. However, in the
  case of regression trees, it is impractical to expand all dimensions,
  which is why we propose our new algorithm. We are following the boosting
  framework, which iteratively fits high-quality decision trees, to
  maximize the objective log-likelihood.

  \subsection{Comparison with Other Consistent List-wise Methods\label{subsec:Comparison-with-Other}}

  Calauzenes et al. \cite{calauzenes2012non} have proven that no consistent
  surrogate function exists for ERR and MAP. However, regarding NDCG,
  Xia et al. \cite{xia2009statistical} proved that the ListMLE model
  is consistent with NDCG@K. They also modified two other losses, cosine
  and KL divergence, to make them NDCG@K consistent. As Xia et al. have
  compared them in their work, we thus compare the PL loss with three
  other consistent versions proposed in \cite{ravikumar2011ndcg}, \emph{squared
  loss}, \emph{cosine}, and \emph{KL divergence}, which were proved
  to be consistent with the whole list, in the case of boosted trees.

  We pay special attention to the first one since it has three different
  implementations. Let $\mathbf{s}$ denote a score vector of all documents,
  $\mathbf{r}$ denote the corresponding relevance vector, and $G(\boldsymbol{r})=2^{\boldsymbol{r}}-1$.
  The consistent and inconsistent equations in terms of square loss
  in \cite{ravikumar2011ndcg} are 
  \begin{equation}
    \phi_{sq}^{consistent}(\mathbf{s},\mathbf{r})=\parallel\mathbf{s}-\frac{G(\mathbf{r})}{\parallel G(\mathbf{r})\parallel_{D}}\parallel_{2}^{2}\label{eq:c-MART-1}
  \end{equation}
  and 
  \begin{equation}
    \phi_{sq}^{inconsistent}(\mathbf{s},\mathbf{r})=\parallel\mathbf{s}-G(\mathbf{r})\parallel_{2}^{2}\label{eq:mart-1}
  \end{equation}
  where the norm $\parallel\cdot\parallel_{D}$ defines the DCG value
  of a ground-truth permutation per query.

  A third equation in \cite{cossock2006subset} is also inconsistent
  with NDCG. 
  \begin{equation}
    \phi_{sq}^{inconsistent}(\mathbf{s},\mathbf{r})=\parallel\mathbf{s}-\mathbf{r}\parallel_{2}^{2}\label{eq:mart-2}
  \end{equation}

  All boosting systems with the least-squares loss are called MART in
  this paper. The two inconsistent versions are point-wise based, and
  the consistent one is list-wise based since the norm $\parallel\cdot\parallel_{D}$
  is operated by query. We remove detailed discussion about the functional
  gradients for all surrogates above due to space limitation.

  \begin{table*}[t]
    \noindent \begin{centering}
    {\footnotesize{}}%
    \begin{tabular}{|c|cl|cccc|cccc|}
      \hline 
      \multicolumn{1}{|c|}{} & {\footnotesize{}System } &  & \multicolumn{4}{c|}{{\footnotesize{}Yahoo 2010}} & \multicolumn{4}{c|}{{\footnotesize{}Microsoft 30K}}\tabularnewline
      \multicolumn{1}{|c|}{} &  &  & {\footnotesize{}NDCG@1 } & {\footnotesize{}@3 } & {\footnotesize{}@10 } & {\footnotesize{}ERR } & {\footnotesize{}NDCG@1 } & {\footnotesize{}@3 } & {\footnotesize{}@10 } & {\footnotesize{}ERR}\tabularnewline
      \hline 
      & {\footnotesize{} } & {\footnotesize{}PLRank(obj=1) } & {\footnotesize{}0.7210 } & {\footnotesize{}0.7267 } & {\footnotesize{}0.7885 } & {\footnotesize{}0.4598 } & {\footnotesize{}0.4947 } & {\footnotesize{}0.4814 } & {\footnotesize{}0.5045 } & {\footnotesize{}0.3770}\tabularnewline
      &  & {\footnotesize{}PLRank(obj=3) } & {\footnotesize{}0.7228 } & {\footnotesize{}0.7290 } & {\footnotesize{}0.7895 } & \textbf{\footnotesize{}0.4611}{\footnotesize{} } & \textbf{\footnotesize{}0.4967}{\footnotesize{} } & \textbf{\footnotesize{}0.4835}{\footnotesize{} } & \textbf{\footnotesize{}0.5063}{\footnotesize{} } & \textbf{\footnotesize{}0.3781}\tabularnewline
      & {\footnotesize{}0} & {\footnotesize{}PLRank(obj=6) } & \textbf{\footnotesize{}0.7240}{\footnotesize{} } & \textbf{\footnotesize{}0.7295}{\footnotesize{} } & \textbf{\footnotesize{}0.7902}{\footnotesize{} } & {\footnotesize{}0.4609 } & {\footnotesize{}0.4949 } & {\footnotesize{}0.4828 } & \textbf{\footnotesize{}0.5069}{\footnotesize{} } & {\footnotesize{}0.3778}\tabularnewline
      \multirow{11}{*}{{\footnotesize{}Trees}} &  & {\footnotesize{}PLRank(obj=9) } & \textbf{\footnotesize{}0.7239}{\footnotesize{} } & \textbf{\footnotesize{}0.7298}{\footnotesize{} } & \textbf{\footnotesize{}0.7903}{\footnotesize{} } & \textbf{\footnotesize{}0.4610}{\footnotesize{} } & {\footnotesize{}- } & {\footnotesize{}- } & {\footnotesize{}- } & {\footnotesize{}-}\tabularnewline
      &  & {\footnotesize{}PLRank(obj=15) } & {\footnotesize{}0.7205 } & {\footnotesize{}0.7291 } & {\footnotesize{}0.7896 } & {\footnotesize{}0.4601 } & {\footnotesize{}- } & {\footnotesize{}- } & {\footnotesize{}- } & {\footnotesize{}-}\tabularnewline
      \cline{2-11} 
      & {\footnotesize{}1 } & {\footnotesize{}LambdaMART } & {\footnotesize{}0.7160 } & {\footnotesize{}0.7187 } & {\footnotesize{}0.7809 } & {\footnotesize{}0.4589 } & {\footnotesize{}0.4942 } & {\footnotesize{}0.4793 } & {\footnotesize{}0.4995 } & \textbf{\footnotesize{}0.3774}\tabularnewline
      \cline{2-11} 
      & {\footnotesize{}2 } & {\footnotesize{}McRank } & {\footnotesize{}0.7213 } & {\footnotesize{}0.7257 } & {\footnotesize{}0.7871 } & {\footnotesize{}0.4586 } & {\footnotesize{}0.4913 } & {\footnotesize{}0.4815 } & \textbf{\footnotesize{}0.5057}{\footnotesize{} } & {\footnotesize{}0.3735}\tabularnewline
      \cline{2-11} 
      & {\footnotesize{}3 } & {\footnotesize{}MART-1 } & {\footnotesize{}0.7112 } & {\footnotesize{}0.7211 } & {\footnotesize{}0.7831 } & {\footnotesize{}0.456 } & {\footnotesize{}0.4856 } & {\footnotesize{}0.4734 } & {\footnotesize{}0.4985 } & {\footnotesize{}0.3769}\tabularnewline
      & {\footnotesize{}4 } & {\footnotesize{}MART-2 } & {\footnotesize{}0.7166 } & {\footnotesize{}0.7230 } & {\footnotesize{}0.7858 } & {\footnotesize{}0.4586 } & {\footnotesize{}0.4924 } & {\footnotesize{}0.4788 } & {\footnotesize{}0.5021 } & {\footnotesize{}0.3736}\tabularnewline
      & {\footnotesize{}ref } & {\footnotesize{}MART in \cite{tyree2011parallel} } & {\footnotesize{}- } & {\footnotesize{}- } & {\footnotesize{}0.782-0.789 } & {\footnotesize{}0.458-0.461 } & {\footnotesize{}- } & {\footnotesize{}- } & {\footnotesize{}- } & {\footnotesize{}-}\tabularnewline
      & {\footnotesize{}5 } & {\footnotesize{}c-MART-1 } & {\footnotesize{}0.7123 } & {\footnotesize{}0.7221 } & {\footnotesize{}0.784 } & {\footnotesize{}0.454 } & {\footnotesize{}0.4860 } & {\footnotesize{}0.4730 } & {\footnotesize{}0.4990 } & {\footnotesize{}0.3750}\tabularnewline
      \cline{2-11} 
      & {\footnotesize{}6 } & {\footnotesize{}CosMART } & {\footnotesize{}0.6979 } & {\footnotesize{}0.6967 } & {\footnotesize{}0.7638 } & {\footnotesize{}0.4521 } & {\footnotesize{}- } & {\footnotesize{}- } & {\footnotesize{}- } & {\footnotesize{}-}\tabularnewline
      & {\footnotesize{}7 } & {\footnotesize{}c-CosMART } & {\footnotesize{}0.6981 } & {\footnotesize{}0.7100 } & {\footnotesize{}0.7669 } & {\footnotesize{}0.4510 } & {\footnotesize{}- } & {\footnotesize{}- } & {\footnotesize{}- } & {\footnotesize{}- }\tabularnewline
      \cline{2-11} 
      & {\footnotesize{}8 } & {\footnotesize{}KLMART } & {\footnotesize{}0.7012 } & {\footnotesize{}0.7111 } & {\footnotesize{}0.7710 } & {\footnotesize{}0.4520 } & {\footnotesize{}- } & {\footnotesize{}- } & {\footnotesize{}- } & {\footnotesize{}- }\tabularnewline
      & {\footnotesize{}9 } & {\footnotesize{}c-KLMART } & {\footnotesize{}0.7020 } & {\footnotesize{}0.7120 } & {\footnotesize{}0.7732 } & {\footnotesize{}0.4525 } & {\footnotesize{}- } & {\footnotesize{}- } & {\footnotesize{}- } & {\footnotesize{}- }\tabularnewline
      \hline 
      \multirow{1}{*}{{\footnotesize{}Linear}} & {\footnotesize{}10 } & {\footnotesize{}CA } & {\footnotesize{}0.6933 } & {\footnotesize{}0.6879 } & {\footnotesize{}0.7549 } & {\footnotesize{}0.444 } & {\footnotesize{}0.4596 } & {\footnotesize{}0.4366 } & {\footnotesize{}0.4597 } & {\footnotesize{}0.3401}\tabularnewline
      \multirow{1}{*}{} & {\footnotesize{}11 } & {\footnotesize{}$l$-ListMLE } & {\footnotesize{}0.7017 } & {\footnotesize{}0.7014 } & {\footnotesize{}0.7673 } & {\footnotesize{}0.4520 } & {\footnotesize{}0.3838 } & {\footnotesize{}0.3880 } & {\footnotesize{}0.4230 } & {\footnotesize{}0.3234}\tabularnewline
      \hline 
    \end{tabular}{\footnotesize\par}
    \par\end{centering}
    \caption{Main results on two real-world datasets. Results on the standard five
    splits of Microsoft data are averaged, and we follow the standard
    one split for the Yahoo data to compare to published results. System
    1 is trained towards optimizing NDCG. System 10, CA, is a Coordinate
    Ascent based method directly maximizing NDCG. We provide results marked
    as ``ref'' reported in other papers. LambdaMART-Aug70 is trained
    on resampled training data, where our experiments are conducted on
    full training data. As PLRank(obj=6) in Microsoft data starts to decrease,
    we did not test more objectives. CosMART got an abnormally low score
    in Microsoft data as $l$-ListMLE, thus it is less meaningful to list
    them. }

    \label{tb:total-results} 
  \end{table*}

  \section{Experiments}

  We studied the performance of the proposed algorithm in two real world
  datasets, Yahoo challenge 2010 and Microsoft 30K. We implemented 9
  baseline ranking systems in C++, which use boosted trees as features.
  System 1 is LambdaMART. System 2 is McRank. System 3 is MART-1 which
  is the first inconsistent version of MART (Eqn. (\ref{eq:mart-1})).
  System 4 is MART-2 which is the second inconsistent version of MART
  (Eqn. $\ref{eq:mart-2}$). System 5 is c-MART-1 which is a consistent
  version of MART-1 (Eqn. (\ref{eq:c-MART-1})). System 6 is CosMART
  which is an inconsistent version of cosine distance loss with boosted
  trees. System 7 is c-CosMART which is a consistent version of CosMART.
  System 8 is KLMART which is a MART using the KL distance. System 9
  is c-KLMART which is a consistent version of KLMART.

  Moreover, in order to compare tree features and linear features, we
  add two linear systems. System 10 is based on a heuristic coordinate
  ascend (CA) based optimization \cite{metzler2007linear} which uses
  linear features and optimizes NDCG directly. CA is used as a reference
  system to represent the average performance of linear systems due
  to its relatively stable and good performances among a variety of
  linear models in different datasets, including the datasets used in
  this work, as shown in the experiments of Tan et al. \cite{tan2013direct}.
  This system is akin to the one proposed by Tan et al., but the latter
  is an exact coordinate ascent optimization ranking method. We also
  used the experimental results in \cite{tan2013direct} as a reference
  here. System 11 is $l$-ListMLE that optimizes top$10$ retrieved
  documents.

  We set up the same parameters as in \cite{wu2010adapting} for all
  systems. The learning rate $\alpha$ is 0.1 (line 15 in Alg. \ref{alg:PL1}).
  We set the number of decision tree leaves as 30, which is a classic
  setting. As in real world datasets, McRank requires more iterations
  to converge, thus we use 2500 boosted trees as a final model, and
  use 1000 boosted trees for other systems. Regarding to PLRank, as
  we mainly concentrate on NDCG@10, we set $K$ to 10 to optimize top$10$
  documents of ground-truth permutations. All results are reported with
  NDCG@(1,3,10) and ERR scores.

  In order to examine the industry-level performance of our system,
  we search exhaustively parameters to compare to the Yahoo Challenge
  results \cite{chapelle2011yahoo} in Table \ref{tbe:ext-yahoo}.

  \subsection{Datasets}

  The LETOR benchmark datasets released in 2007 \cite{qin2010letor}
  have significantly boosted the development of learning to rank algorithms
  since researchers could compare their algorithms on the same datasets
  for the first time. But unfortunately, the sizes of the datasets in
  LETOR are several orders of magnitude smaller than the ones used by
  search engine companies. Several researchers have noticed that the
  conclusions drawn from experiments based on LETOR datasets are unstable
  and quite different from the ones based on large real datasets \cite{chapelle2011yahoo}.
  Thus in this work, we attempt to make stable system comparisons by
  using as large datasets as possible, and we use two real world datasets,
  Yahoo challenge 2010 and Microsoft 30K. The statistics oh these three
  data sets are reported in Table \ref{tb:dataset} which might a bit
  different from those in \cite{chapelle2011yahoo} as we only give
  the statistics of training datasets.

  \begin{table}[h]
    \noindent \begin{centering}
    {\small{}}%
    \begin{tabular}{c|cccc}
      & {\small{}\#Query } & {\small{}\#Doc. } & {\small{}\#D. / \#Q. } & {\small{}\#Feat.}\tabularnewline
      \hline 
      {\small{}Microsoft 30K} & {\small{}18.9K } & {\small{}2270K } & {\small{}120 } & {\small{}136}\tabularnewline
      {\small{}Yahoo 2010 (Set 1) } & {\small{}20K } & {\small{}473K } & {\small{}23 } & {\small{}519}\tabularnewline
      \hline 
      {\small{}McRank\cite{li2007mcrank} } & {\small{}10-26K } & {\small{}474-1741K } & {\small{}18-88 } & {\small{}367-619}\tabularnewline
      {\small{}LambdaMART\cite{wu2010adapting} } & {\small{}31K } & {\small{}4154K } & {\small{}134 } & {\small{}416}\tabularnewline
      {\small{}Ohsumed } & {\small{}106 } & {\small{}16K } & {\small{}150 } & {\small{}45}\tabularnewline
      {\small{}LETOR 4.0 } & {\small{}2.4K } & {\small{}85K } & {\small{}34 } & {\small{}46}\tabularnewline
    \end{tabular}{\small\par}
    \par\end{centering}
    \caption{The top two datasets are used, while the others are as a reference.
    Ohsumed is of LETOR 3.0. \#D./\#Q. means average document number per
    query. }

    \label{tb:dataset} 
  \end{table}
  Microsoft 30K is the largest publicly released dataset in terms of
  the document number. As its official release has provided a standard
  5-fold split, we report average results. Regarding the Yahoo dataset,
  it only provides a 1-fold split. In order to compare to other released
  systems, we report results on the standard 1-fold split in Table \ref{tb:total-results},
  and report average results on a randomly generated 3-fold split in
  Figure \ref{fig:main-tree-comp}. 

  \subsection{$l$-ListMLE vs. Other Linear Systems}

\begin{figure}[h]
    \noindent \begin{centering}
    \includegraphics[scale=0.4]{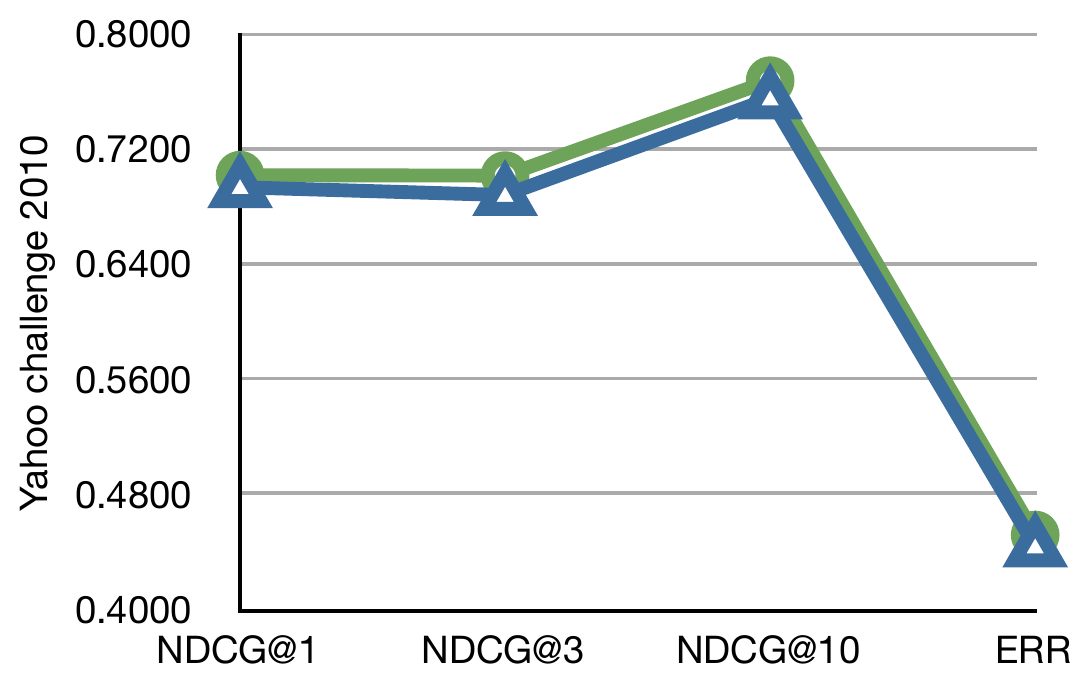}\includegraphics[scale=0.4]{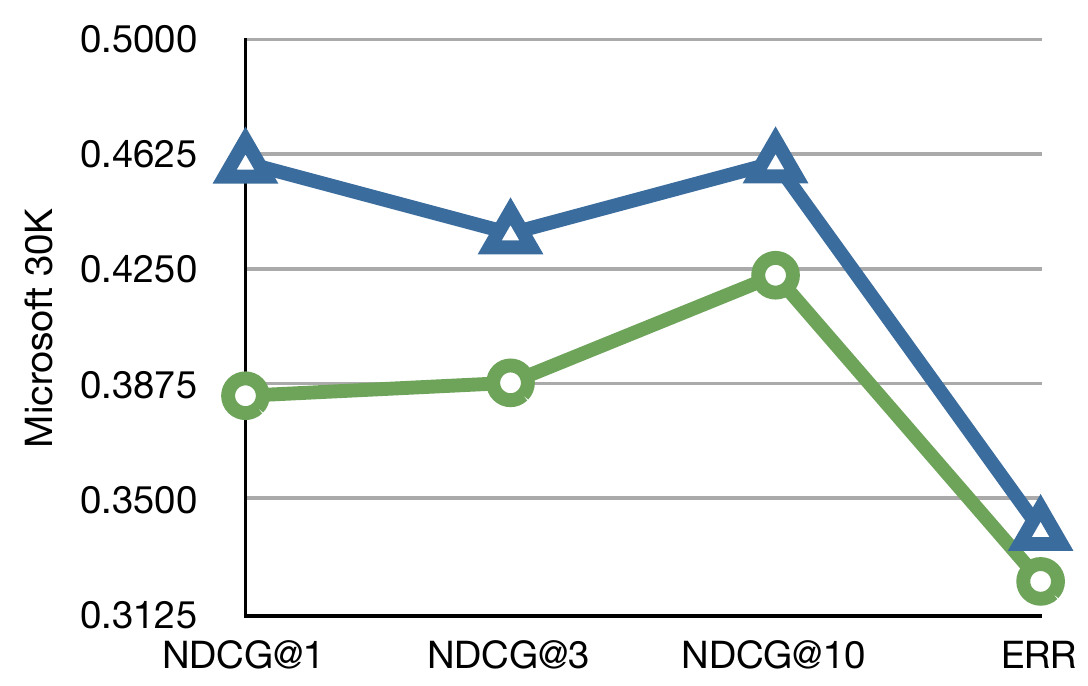} 
    \par\end{centering}
    \noindent \begin{centering}
    \includegraphics[scale=0.5]{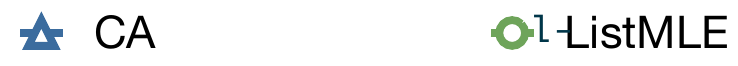} 
    \par\end{centering}
    \caption{The performance of $l$-ListMLE and the selected linear reference
    system Coordinate Ascent (CA) on Yahoo data (left) and Microsoft 30K
    (right). CA is capable of representing the mainstream linear systems
    on these datasets \cite{tan2013direct}. }

    \label{fig:stable-com} 
  \end{figure}

  We first examine the performance of $l$-ListMLE (System-11) compared
  to another linear system CA (System-10). Their results are shown in
  Table \ref{tb:total-results} and Figure \ref{fig:stable-com}. $l$-ListMLE
  obtains 0.7673 in NDCG@10 in the Yahoo 2010 dataset after 100 iterations
  of quasi-Newton optimization, but performs unsatisfactorily in Microsoft
  30K even after 1000 iterations, approximately 8 percent lower in NDCG@1,
  and several percent lower in other measures. Tan et al. \cite{tan2013direct}
  also compared several linear systems in these two datasets, except
  $l$-ListMLE. Our implementation of $l$-ListMLE outperforms their
  best result 0.760 from DirectRank in the Yahoo datasets, while performs
  significant worse in the Microsoft 30K. 

\begin{figure}[h]
    \noindent \begin{centering}
    \includegraphics[scale=0.4]{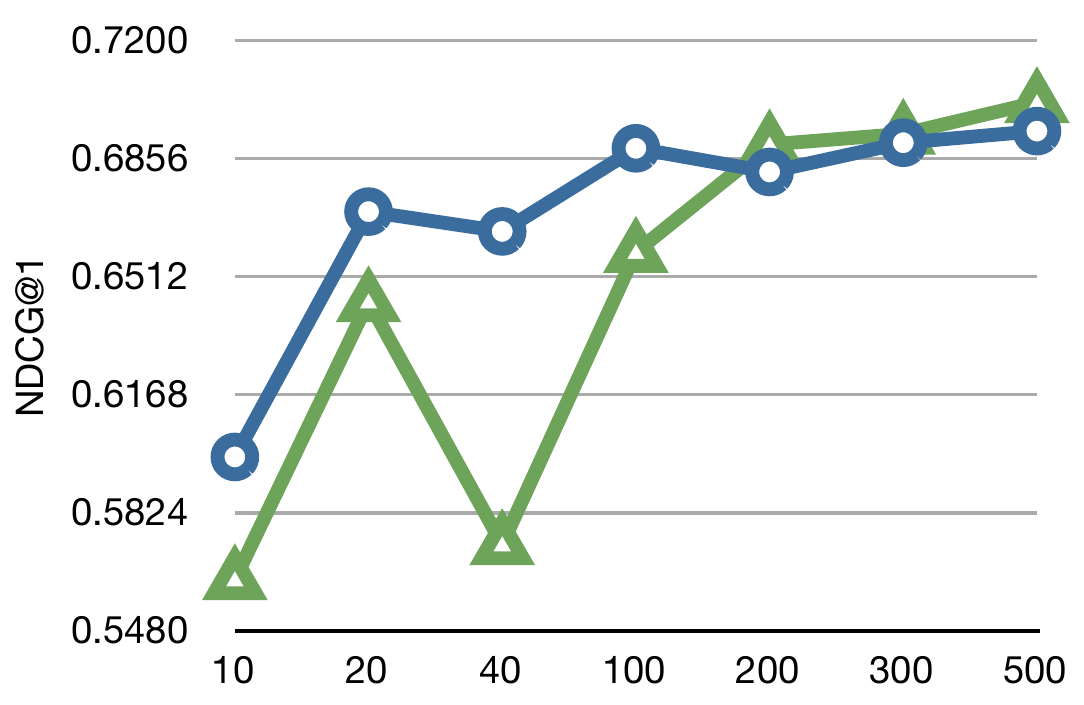}\includegraphics[scale=0.4]{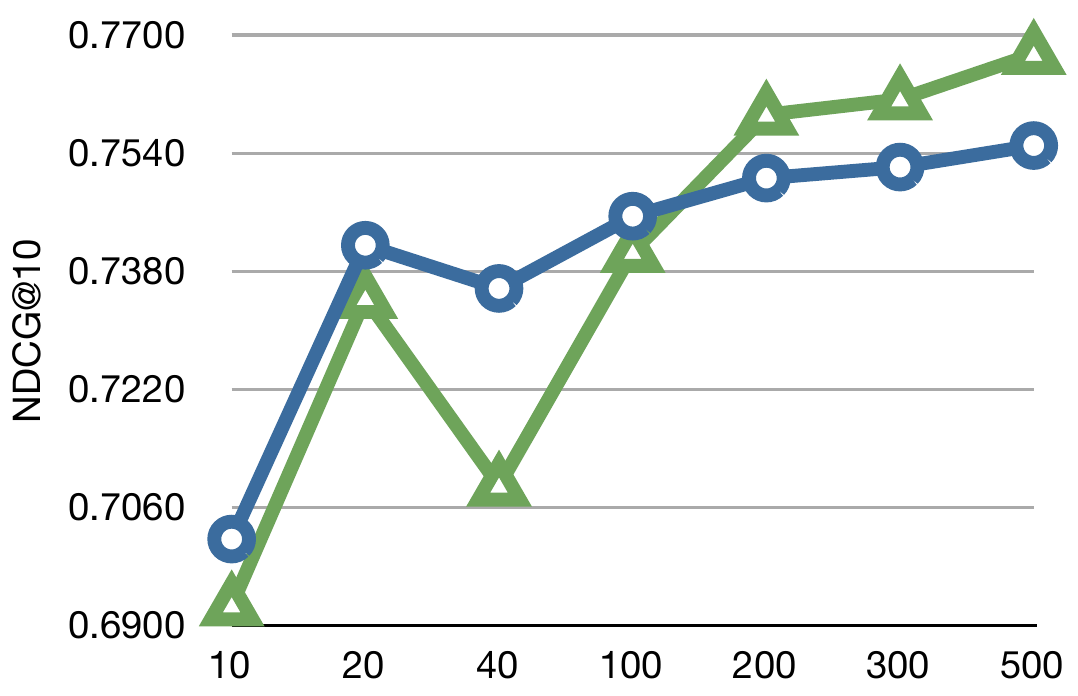} 
    \par\end{centering}
    \noindent \begin{centering}
    \includegraphics[scale=0.5]{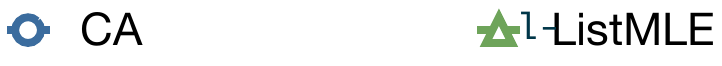} 
    \par\end{centering}
    \caption{l-ListMLE and CA with different number of features.}

    \label{fig:vary-feat-yahoo1} 
  \end{figure}

  The unexpectedly bad performance of ListMLE in the larger dataset
  contradicts the proof from \cite{xia2009statistical}, that is ListMLE
  is consistent with NDCG. In another words, ListMLE theoretically should
  perform better with more available data. The main reason may be that
  the features on Microsoft 30K is not rich enough to ensure the consistency
  of ListMLE. To verify this, we notice that the features of Yahoo 2010
  data set are richer than Microsoft 30k, thus we conduct experiments
  on Yahoo 2010 dataset by adjusting the number of features and compare
  the performance of $l$-ListMLE and CA. The results are shown in Figure
  \ref{fig:vary-feat-yahoo1}. Since the features might not be independent
  to each other, the NDCG performance curves are not monotonic with
  the size of features number. However both figures have their own critical
  points, 200 for NDCG@1 and 100 for NDCG@10: When the feature number
  is beyond this point, $l$-ListMLE beats CA, otherwise it performs
  worse than CA.

  To improve the performance of $l$-ListMLE, instead of using a linear
  feature model, we need to increase the model capacity that have more
  expressive power. Thus we decide to use decision trees as our basic
  weak learners, and we grown our model through gradient boosting that
  maximize the likelihood of ground-truth ranks. The PL loss is not
  the only one that is consistent with NDCG, there are other three models
  proposed in \cite{ravikumar2011ndcg} that are also consistent with
  it, so we extend these three models to boosted trees versions for
  a full comparison.

  \subsection{Different Number of Ground Truth Permutations}

  \begin{table}[h]
    \noindent \begin{centering}
    \begin{tabular}{c|cc|cc}
      {\small{}\#Obj } & \multicolumn{2}{c|}{{\small{}Yahoo}} & \multicolumn{2}{c}{{\small{}Microsoft}}\tabularnewline
      \hline 
      {\small{}3 } & {\small{}0.8672 } & {\small{}2.60 } & {\small{}0.8986 } & {\small{}2.70}\tabularnewline
      {\small{}6 } & {\small{}0.7586 } & {\small{}4.5 } & {\small{}0.8069 } & {\small{}4.84}\tabularnewline
      {\small{}9 } & {\small{}0.6965 } & {\small{}6.27 } & {\small{}0.7609 } & {\small{}6.85}\tabularnewline
      {\small{}12 } & {\small{}0.6524 } & {\small{}7.83 } & {\small{}0.7277 } & {\small{}8.73}\tabularnewline
      {\small{}15 } & {\small{}0.6197 } & {\small{}9.28 } & {\small{}0.7026 } & {\small{}10.54}\tabularnewline
      {\small{}18 } & {\small{}0.5926 } & {\small{}10.66 } & {\small{}0.6823 } & {\small{}12.28}\tabularnewline
      {\small{}21 } & {\small{}0.5708 } & {\small{}11.97 } & {\small{}0.6661 } & {\small{}13.99}\tabularnewline
    \end{tabular}
    \par\end{centering}
    \caption{Compression ratio of denominator terms in Eqn. (\ref{eq: full-formula})
    in considering multiple ground-truth permutations.}

    \label{tbe:balance} 
  \end{table}
  We empirically search an optimal setting to balance the running time
  and performance.Table \ref{tbe:balance} displays actual compression
  ratio. For example, when objective number is 9, actual number of terms
  in computing the functional gradient is 69.6 percent of that without
  compressed storage, and this is equivalent to 6.27 objectives. From
  the results in Tables \ref{tb:total-results}, \ref{tbe:run-time},
  \ref{tbe:balance}, we recommend to use PL(obj=3) in practice to gain
  stable improvements with acceptable extra training time.

  \subsection{PLRank vs. $l$-ListMLE, MART, McRank and LambdaMART}

  \begin{figure}[h]
    \noindent \begin{centering}
    \includegraphics[scale=0.425]{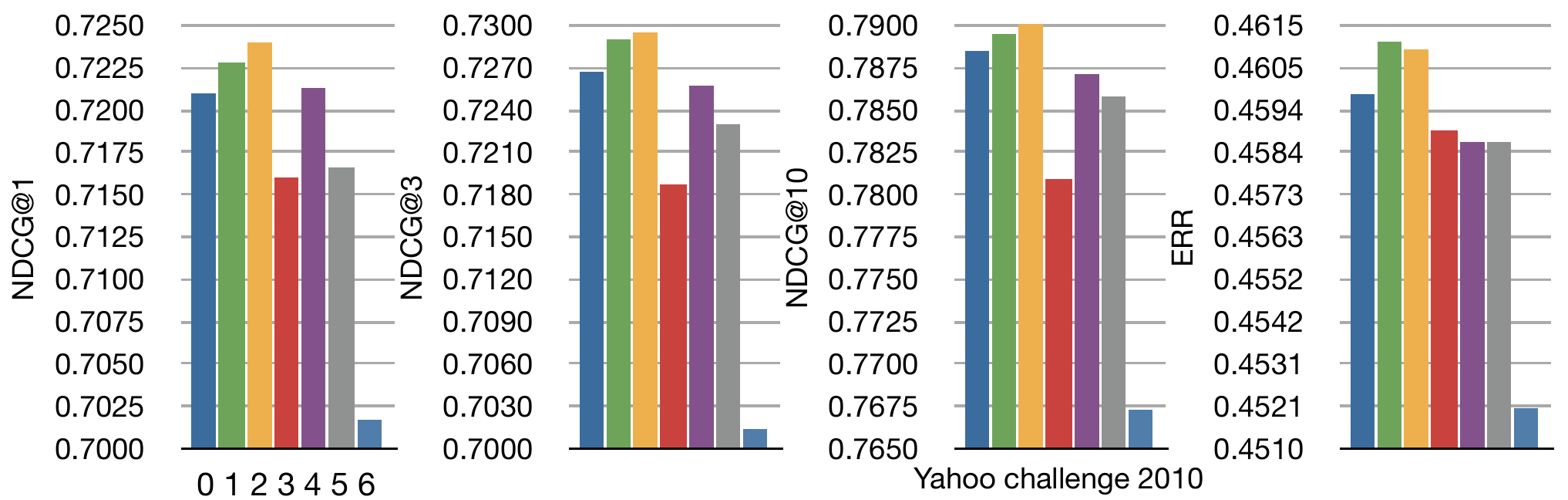} 
    \par\end{centering}
    \noindent \begin{centering}
    \includegraphics[scale=0.425]{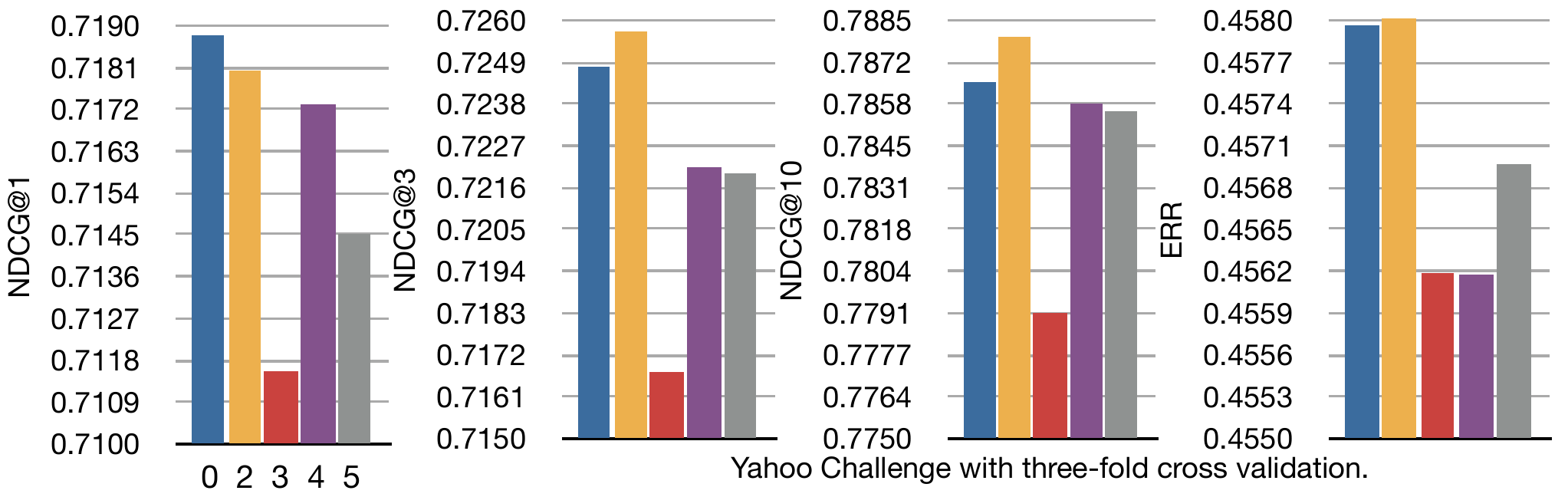} 
    \par\end{centering}
    \noindent \begin{centering}
    \includegraphics[scale=0.425]{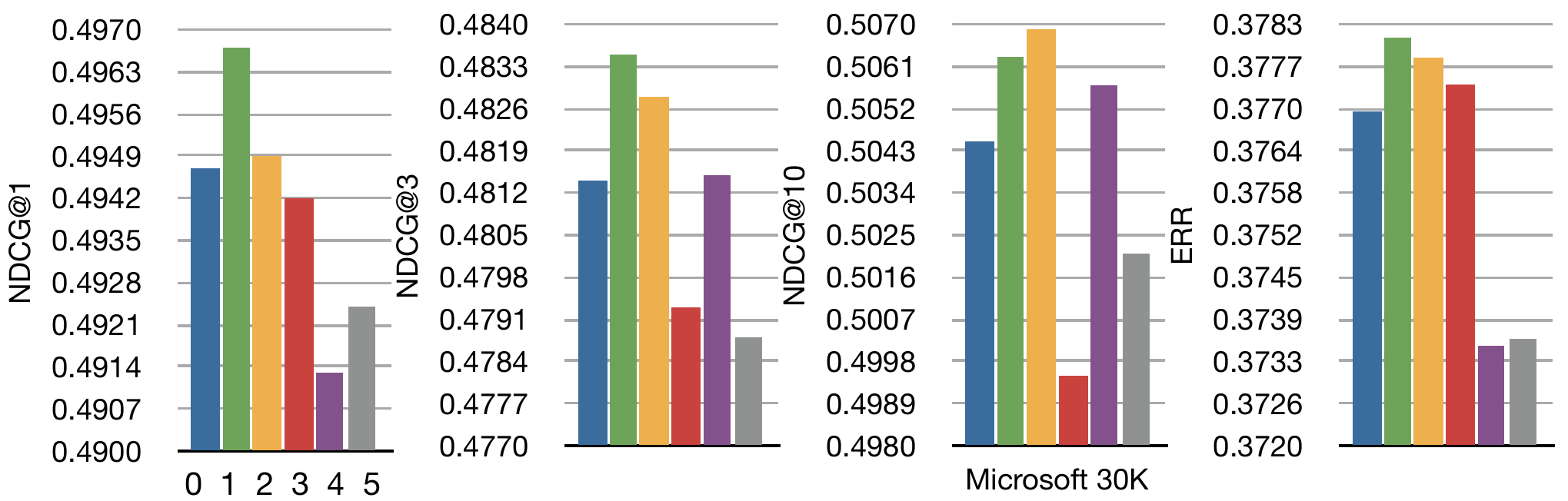} 
    \par\end{centering}
    \noindent \begin{centering}
    \includegraphics[scale=0.5]{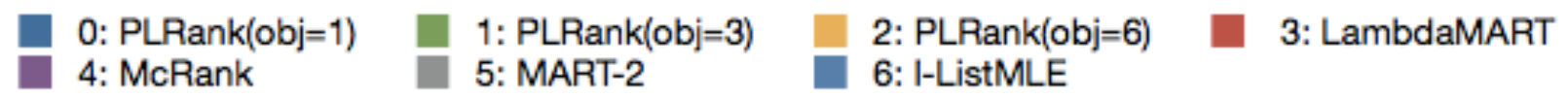} 
    \par\end{centering}
    \caption{Comparison of several tree-based systems. }

    \label{fig:main-tree-comp} 
  \end{figure}
  Currently, the state-of-the-art learning to rank systems use boosted
  trees which have been proven to be more powerful than those using
  linear features in real world datasets. The champion of Yahoo challenge
  2010 is a system that combines approximately 12 models, most of which
  are trained with LambdaMART \cite{burges2011learning}. The other
  two state-of-the-art systems using trees are MART and McRank, one
  optimizes least-square loss and the other treat the ranking as a multi-class
  classification.

  As shown in Table \ref{tb:total-results}, PLRank outperforms $l$-ListMLE,
  which is a natural result as PLRank is in a more complex function
  space than the linear space. However, what surprises us is that, in
  the Yahoo dataset there are moderate improvements, approximately 2
  points in NDCG(@1, 3, 10), while in the Microsoft dataset, there are
  significant 8 to 10 points in NDCG(@1, 3, 10). On one aspect, boosted
  trees indeed could capture the dependency between features, and on
  another aspect, it is especially effective for the PL loss when the
  features are not rich.

  As shown in Figure \ref{fig:main-tree-comp} and Table \ref{tb:total-results},
  the tree-based systems obviously perform well over linear feature
  systems. Among tree-based systems, PLRank demonstrates some moderate
  improvements over MART, McRank and LambdaMART in the Yahoo dataset,
  and in the Microsoft dataset, all tree-based systems perform pretty
  closely to each other.

  McRank and PLRank are more close in six NDCG scores except NDCG@1
  in the Microsoft dataset. LambdaMART performs well in ERR, and is
  significantly better than McRank and MART, and close to PLRank(obj=1).
  Comparatively, three PLRank variants act more stably. PLRank(obj=1)
  is always in best two systems on all measures when it is compared
  with McRank, on the other hand, as shown in Table 1 LambdaMART and
  MART. PLRank(obj=2) is considered to be the best in balancing the
  performance and running time.

  Two-tailed t-test results show PLRank(obj={*}) systems would have
  significant improvements over others when their differences are greater
  than about 0.5 point at 95\% confidence. Unfortunately, in Table \ref{tb:total-results},
  most of the improvements of PLRank(obj={*}) are not significant, just
  matchable to these state-of-the-art systems.

  Our MART baseline results are close to those reported in \cite{tyree2011parallel}.
  Tan et al. \cite{tan2013direct} also used the same datasets to compare
  LambdaMART and MART, and their baselines are about 1 point lower in
  NDCG than our reported results. We notice that their baselines are
  from RankLib, which is written in Java, and DirectRank is implemented
  in C++. In comparison, our 10 tree-based systems are re-implemented
  in C++ with an identical code template, thus our systems could be
  better to reflect differences in models rather than being impacted
  by coding. 

  \subsubsection{PLRank vs.Other Consistent List-wise Method with Boosted regression
  trees}

  The list-wise methods discussed in Section \ref{subsec:Comparison-with-Other}
  have better performance than their in-consistent counterparts in Yahoo
  dataset, although the differences are not that much. In contrast,
  it is reported in \cite{ravikumar2011ndcg} that for all linear systems,
  the consistent versions improves NDCG scores of the in-consistent
  counterparts by several points.

  As shown in Table \ref{tb:total-results}, these consistent methods,
  after extended to boosted trees versions, unfortunately, have not
  show competitive performances when compared with LambdaMART, McRank
  and PLRank, so we did not run them on the larger Microsoft dataset.

  LambdaMART is a method that considers NDCG loss in optimization, and
  McRank optimizes unnormalized NDCG, so we only need to further analyze
  the surrogate functions of PLRank and the three consistent versions,
  that are not directly related to NDCG. A plausible explanation is
  the PL loss is consistent with NDCG@$K$, $K$ taken 10, while those
  of c-MART-1, c-CosMART, c-KLMART are consistent with NDCG with a whole
  list. We conjecture that when we let $K$ go to the whole list, these
  systems would show advantages.

  \subsubsection{Running time}

  \begin{table}[h]
    \noindent \begin{centering}
    {\small{}}%
    \begin{tabular}{c|cccccc}
      & {\small{}MT } & {\small{}PR1 } & {\small{}LM } & {\small{}PR3 } & {\small{}PR6 } & {\small{}MR }\tabularnewline
      \hline 
      {\small{}Hours } & {\small{}83 } & {\small{}91 } & {\small{}101 } & {\small{}106.2 } & {\small{}126 } & {\small{}250+ }\tabularnewline
    \end{tabular}{\small\par}
    \par\end{centering}
    \caption{ Sorted running time in the Microsoft dataset in a single computing
    core. MT: MART. PR$n$: PLRank(obj=$n$). MR: McRank. }

    \label{tbe:run-time} 
  \end{table}
  The computational costs of tree-based systems are mainly at the stage
  of tree construction, thus these systems have the same time complexity
  except that McRank requires more iterations to reach reasonable performance.
  The running times of PLRank, MART, LambdaMART and McRank in Microsoft
  dataset are shown in Table \ref{tbe:run-time}. Their differences
  are mainly due to the computation of functional gradients.

  \subsection{Industry-level Comparison}

  Last, in Table \ref{tbe:ext-yahoo}, we examine our PLRank system
  in the Yahoo Challenge set 1 data in an industry level. To save time,
  we use PLRank(obj=1) and search its parameters to gain best performance
  regardless of any cost. We sweep the number of tree leaves from 100
  to 1000 in steps of 100, and the learning rate $\alpha$ from 0.01
  to 0.1 in steps of 0.03. We notice that \cite{burges2011learning}
  actually did not release results of single LambdaMART systems in the
  standard test set, but in a self-define test set. Since the final
  result of a LambdaMART-based system combination in the standard set
  has been available, we reasonably estimate their single LambdaMART
  systems in the standard test set.

  \begin{table}[h]
    \noindent \begin{centering}
    {\small{}}%
    \begin{tabular}{c|cc}
      {\small{}\#System } & \multicolumn{1}{c}{{\small{}NDCG@10}} & \multicolumn{1}{c}{{\small{}ERR}}\tabularnewline
      \hline 
      {\small{}PLRank(obj=1) } & {\small{}0.802} & {\small{}0.4660}\tabularnewline
      {\small{}LambdaMART in \cite{burges2011learning}} & {\small{}0.796} & {\small{}0.4649}\tabularnewline
      {\small{}LambdaMART-Aug70 in \cite{burges2011learning}} & {\small{}0.804} & {\small{}0.4669}\tabularnewline
    \end{tabular}{\small\par}
    \par\end{centering}
    \caption{An industry level comparison in the standard Yahoo Challenge set 1
    data with LambdaMART.}

    \label{tbe:ext-yahoo} 
  \end{table}
  LambdaMART with complete training set for tuning parameters reaches
  0.796 in NDCG@10, and they use a resampled technique called Aug70
  to increase the training data to improve their systems to 0.804. In
  comparison, our result is acceptable compared with LambdaMART with
  standard training data for tuning, as our result is obtained in a
  resources-constrained laboratory environment, which might be better
  given industry-level computing clusters for larger parameter searching. 

  \section{Conclusion}

  As a non-linear algorithm in the boosting framework, our proposed
  PLRank enriches the ListMLE framework. As far as we know, PLRank is
  the first list-wise based ranking system that in real-world datasets
  could match or outperform suitably the famous LambdaMART and McRank
  in terms of NDCG and ERR. 

  \bibliographystyle{plain}
  \nocite{*}
  \bibliography{references}
\end{document}